\documentstyle[12pt]{article}

\newlength{\titlesep}
\makeatletter

\renewcommand{\thesection}{\Roman{section}}
\@addtoreset{equation}{section}
\renewcommand{\theequation}{\arabic{section}.\arabic{equation}}
\renewcommand{\appendix}{\par
  \setcounter{section}{0}
  \setcounter{subsection}{0}
  \renewcommand{\thesection}{Appendix~\Alph{section}}
  \renewcommand{\theequation}{\Alph{section}.\arabic{equation}}}

\def\fnum@figure{FIG.~\thefigure}
\newcommand{\Fig}[1]{\item[{\bf FIG.~\protect\ref{#1}}]}

\newcounter{figureparent}
\@addtoreset{figure}{figureparent}

\newcounter{eqnparent}
\@addtoreset{equation}{eqnparent}

\renewcommand{\abstract}{\if@twocolumn
  \section*{Abstract}
  \else
  \begin{center}
    {\bf Abstract\vspace{-.5em}\vspace{0pt}}
  \end{center}
  \quotation
  \fi}
\renewcommand{\endabstract}{\if@twocolumn\else\endquotation\fi}

\newcommand{\preprintnumber}[1]
{\begin{flushright}
  \begin{tabular}{l} #1 \end{tabular}
  \end{flushright}}

\makeatother

\newcommand{\gsim}%
{\mbox{\raisebox{-1.0ex}
    {$\ \stackrel{\textstyle >}{\textstyle \sim}\ $}}}
\newcommand{\lsim}%
 {\mbox{\raisebox{-1.0ex}
     {$\ \stackrel{\textstyle <}{\textstyle \sim}\ $}}}
\newcommand{\ie}{{\it i.e.\ }}
\newcommand{\etc}{{\it etc.\ }}
\newcommand{\etal}{{\it et al.\ }}
\newcommand{\vev}[1]{\left\langle #1 \right\rangle}

\newcommand{\str}{\ {\rm Str}\ }
\newcommand{\repr}[1]{{\boldmath$#1$}}
\newcommand{\ol}[1]{\overline{#1}}
\newcommand{\MeV}{\ {\rm MeV}\ }
\newcommand{\GeV}{\ {\rm GeV}\ }
\newcommand{\TeV}{\ {\rm TeV}\ }

\newcommand{\Journal}[4]{{#1} {\bf #2}, {#4} {(#3)}}
\newcommand{\pl}{\sl Phys.~Lett.}
\newcommand{\prp}{\sl Phys.~Rep.}
\newcommand{\pr}{\sl Phys.~Rev.}
\newcommand{\prl}{\sl Phys.~Rev.~Lett.}
\newcommand{\np}{\sl Nucl.~Phys.}
\newcommand{\ptp}{\sl Prog.~Theor.~Phys.}

\newcommand{\ibid}{\it ibid.}

\begin{document}
\baselineskip 18pt

\begin{titlepage}

\preprintnumber{ICRR-Report-317-94-12 \\ April 1994}
\vspace*{\titlesep}
\begin{center}
{\LARGE\bf Flavor mixing in the gluino coupling \\
  and the nucleon decay}\\
\vspace*{\titlesep}
{\large Toru {\sc Goto}, Takeshi {\sc Nihei}} and
{\large Jiro {\sc Arafune}}\\
\vspace*{\titlesep}
{\it Institute for Cosmic Ray Research, University of Tokyo,\\
  Midori-cho, Tanashi-shi, Tokyo 188 JAPAN}\\
\end{center}
\vspace*{\titlesep}
\begin{abstract}
Flavor mixing in the quark-squark-gluino coupling is studied for the
minimal SU(5) SUGRA-GUT model and applied to evaluation of the nucleon
lifetime.
All off-diagonal (generation mixing) elements of Yukawa coupling
matrices and of squark/slepton mass matrices are included in solving
numerically one-loop renormalization group equations for MSSM
parameters, and the parameter region consistent with the radiative
electroweak symmetry breaking condition is searched.
It is shown that the flavor mixing in the gluino coupling for a large
$\tan\beta$ is of the same order of magnitude as the corresponding
Kobayashi-Maskawa matrix element in both up-type and down-type sector.
There exist parameter regions where the nucleon decay amplitudes for
charged lepton modes are dominated by the gluino dressing process,
while for all the examined regions the neutrino mode amplitudes are
dominated by the wino dressing over the gluino dressing.
\end{abstract}

\end{titlepage}


\section{Introduction}
\label{sec:introduction}

SU(5) supersymmetric (SUSY) grand unified theory (GUT) is an
attractive candidate for the unified theory of strong and electroweak
interactions.
The analyses of the gauge coupling unification \cite{LL-ABF-M} suggest
the validity of the minimal supersymmetric standard model (MSSM)
just above the electroweak scale $\sim 100 \GeV$ and the unification
of SU(3)$\times$SU(2)$\times$U(1) gauge group into a simple SU(5) at
the GUT scale $M_X \sim 10^{16} \GeV$.

One of the features of MSSM/SUSY-GUT is the existence of soft SUSY
breaking.
It gives quarks (or leptons) and their superpartners different mass
matrices in the generation (flavor) space.
In results, due to the discrepancy between the mass diagonalizing
bases of quarks and those of squarks, a generation mixing occurs in
the quark-squark-gaugino coupling (gluino coupling, in particular),
which may give considerable contributions to the nucleon decay, flavor
changing neutral currents (FCNC) and other various phenomena in
SUSY-GUT.
The flavor mixing in the gaugino coupling plays an important role in
the nucleon decay, since its amplitude is dominated by the dimension
five interaction followed by the gaugino ``dressing'' process
\cite{SY-W} in the minimal SU(5) SUSY-GUT model.
However, only a simplified treatment is made in the previous analyses
of the nucleon decay \cite{HMY,MATS,NCA,ENR}, where the diagonal
(in the generation space) gluino coupling is assumed leading to the
negligible contribution from the gluino dressing process.

The purpose of the present paper is to study the flavor mixing in the
gaugino coupling extensively and evaluate the contribution of the
gluino dressing process to the nucleon decay amplitude.
Based on the minimal SU(5) supergravity (SUGRA-) GUT \cite{N-LN},
we assume that the soft SUSY breaking parameters are ``universal'' at
the GUT scale \cite{BFS-CAN-HLW}.
We include all off-diagonal elements of Yukawa coupling matrices and
of squark/slepton mass matrices in solving numerically one-loop
renormalization group equations (RGEs)%
\footnote{Two-loop RGEs for soft SUSY breaking parameters are obtained
  recently \protect\cite{Y-MH}, which will be important for more
  accurate analysis.}
for all MSSM parameters \cite{IKKT-APW,BKS} with the universal
boundary conditions.
We then evaluate the effective potential for the Higgs fields at the
electroweak scale to find a consistent SU(2)$\times$U(1) breaking
minimum in accordance with the radiative electroweak symmetry breaking
scenario \cite{IKKT-APW}.
The obtained mass matrices of all particles are diagonalized to
evaluate the flavor mixing in the gaugino couplings.
The mass spectrum and the mixing are then used to calculate the
nucleon decay amplitudes for various decay modes.
It is found that the flavor mixing in the gluino coupling depends on
$\tan\beta$ (ratio of vacuum expectation values of the Higgs doublets)
and is roughly of the same order of magnitude as the corresponding
Kobayashi-Maskawa matrix element.
As for the nucleon decay, we find the gluino dressing process
dominates the amplitude for the decay modes containing a charged
lepton (and a meson) if $\tan\beta$ is large and if the gluino mass is
much smaller than the squark masses.
Note that the flavor mixing in the gaugino couplings is studied
previously in a systematic analysis of FCNC \cite{BKS} with a
semi-analytic solution of the RGEs for small $\tan\beta$ (top Yukawa
coupling $\gg$ bottom Yukawa coupling).
On the contrary, our numerical method cover the whole range of
$\tan\beta$, since all Yukawa couplings are taken into account.

The remaining part of this paper is organized as follows.
After introducing in the next section the minimal SU(5) SUGRA-GUT
model which we consider, we formulate the flavor mixing in the gaugino
couplings in Sec.~\ref{sec:mixing}.
In Sec.~\ref{sec:decay} the nucleon decay amplitudes are obtained with
a careful treatment of the flavor mixing.
The outline of the numerical calculation and the results are presented
in Sec.~\ref{sec:calculations} and our conclusions are summarized in
Sec.~\ref{sec:conclusion}.

\section{Minimal SU(5) SUGRA-GUT model}
\label{sec:model}


Minimal SU(5) SUSY-GUT model contains three generations of matter
multiplets with \repr{10} and \repr{\ol{5}} representations,
$\Psi_i^{AB}$ and $\Phi_{jA}$ respectively, where suffices
$A,B=1,2,\cdots,5$ are SU(5) indices and $i,j=1,2,3$ are the
generation labels, and three kinds of Higgs multiplets with \repr{5},
\repr{\ol{5}} and \repr{24} representations, $H_5^A$, $H_{\ol{5}A}$
and $\Sigma^A_B$ respectively.
SU(5) and R-parity invariant superpotential $W_{\rm GUT}$ at the GUT
scale is written as
\begin{eqnarray}
  W_{\rm GUT}(M_X) &=& f^{ij} \Psi_i^{AB} \Phi_{jA} H_{\ol{5}B}
        + \frac{1}{8} g^{ij} \epsilon_{ABCDE}
          \Psi_i^{AB} \Psi_j^{CD} H_5^E
\nonumber\\
    & & + \lambda H_{\ol{5}A}
          \left( \Sigma^A_B + M \delta^A_B \right) H_5^B
        + W_\Sigma \left( \Sigma \right) ~.
  \label{GUTsuperpotential}
\end{eqnarray}
Here, $\epsilon_{ABCDE}$ is the totally antisymmetric constant,
$f^{ij}$, $g^{ij}=g^{ji}$ and $\lambda$ are dimensionless couplings,
$M$ is a mass parameter and $W_\Sigma$ is a self-interaction
superpotential for $\Sigma^A_B$.
SU(5) symmetry is spontaneously broken down to SU(3) $\times$ SU(2)
$\times$ U(1) with the nonvanishing vacuum expectation value of the
adjoint Higgs $\Sigma^A_B$.
Below the GUT scale, the model is reduced to MSSM with effective
higher dimensional operators, which are obtained by integrating out
the superheavy particles in Eq.~(\ref{GUTsuperpotential}).
The effective superpotential is then written as
\begin{eqnarray}
  W_{\rm eff} &=& W_{\rm MSSM} + W_5 + O(M_X^{-2}) ~,
  \nonumber\\
  W_{\rm MSSM} &=& f_D^{ij} Q_i^{a\alpha} D_{ja} H_{1\alpha}
      + f_L^{ij} \epsilon^{\alpha\beta} E_i L_{j\alpha} H_{1\beta}
      + g_U^{ij} \epsilon_{\alpha\beta} Q_i^{a\alpha} U_{ja} H_2^\beta
      + \mu H_{1\alpha} H_2^\alpha ~,
  \nonumber\\
  W_5 &=& \frac{1}{M_C} \left\{
        \frac{1}{2} C_L^{ijkl} \epsilon_{abc} \epsilon_{\alpha\beta}
               Q_k^{a\alpha} Q_l^{b\beta} Q_i^{c\gamma} L_{j\gamma}
      + C_R^{ijkl} \epsilon^{abc} U_{ia} D_{jc} E_k U_{lb} \right.
  \nonumber\\
  & & ~~~~~~~+ ( \mbox{baryon number/lepton number
                       conserving terms} ) \biggr\} ~.
  \label{superpotential}
\end{eqnarray}
Here, $Q$, $U$ and $E$ are chiral superfields which contain
left-handed quark doublet, right-handed up-type quark and right-handed
charged lepton respectively, and are embedded in $\Psi$ (to be
specified in Eq.~(\ref{Embedding}));
 $D$ and $L$, which are embedded in $\Phi$, contain right-handed
 down-type quark and left-handed lepton doublet respectively;
$H_1$ and $H_2$ are Higgs doublets embedded in $H_{\ol{5}}$ and $H_5$,
respectively.
The suffices $a,b,c=1,2,3$ are SU(3) indices and $\alpha,\beta=1,2$
are SU(2) indices.
$M_C$ is the colored Higgs mass which is assumed to be $O(M_X)$, while
the supersymmetric mass of Higgs doublet $\mu$ is of the order of the
$Z$ boson mass $m_Z$.
This discrepancy is owing to a tree level fine-tuning in the GUT
superpotential.
At the GUT scale, $f_D$ and $f_L$ are unified and $C_L$ and $C_R$ are
written in terms of the Yukawa coupling constants (see
Sec.~\ref{sec:decay}).
Baryon number (and lepton number) violating terms in $W_5$ give
dominant contributions to the nucleon decay in this model.


In addition to the supersymmetric Lagrangian to be derived from
(\ref{superpotential}), the following soft SUSY breaking terms are
included:
\begin{eqnarray}
  -{\cal L}_{\rm soft} &=&
       \left(m_Q^2\right)_i^j \tilde{q}^{\dagger i} \tilde{q}_j
     + \left(m_D^2\right)_i^j \tilde{d}^{\dagger i} \tilde{d}_j
     + \left(m_U^2\right)_i^j \tilde{u}^{\dagger i} \tilde{u}_j
  \nonumber\\
  && + \left(m_L^2\right)_i^j \tilde{l}^{\dagger i} \tilde{l}_j
     + \left(m_E^2\right)_i^j \tilde{e}^{\dagger i} \tilde{e}_j
     + \Delta_1^2 h_1^\dagger h_1 + \Delta_2^2 h_2^\dagger h_2
  \nonumber\\
  && + \left\{   A_D^{ij} \tilde{q}_i \tilde{d}_j h_1
               + A_L^{ij} \tilde{e}_i \tilde{l}_j h_1
               + A_U^{ij} \tilde{q}_i \tilde{u}_j h_2
               - B\mu h_1 h_2 + {\rm h.~c.} \right\}
  \nonumber\\
  && + \frac{1}{2} \left\{   M_1 \tilde{B} \tilde{B}
                           + M_2 \tilde{W} \tilde{W}
                           + M_3 \tilde{G} \tilde{G}
                           + {\rm h.~c.} \right\} ~,
  \label{softbreaking}
\end{eqnarray}
where $\tilde{q}_i$, $\tilde{d}_i$, $\tilde{u}_i$, $\tilde{e}_i$,
$\tilde{l}_i$, $h_1$ and $h_2$ are scalar components of $Q_i$, $D_i$,
$U_i$, $E_i$, $L_i$, $H_1$ and $H_2$, respectively, and $\tilde{B}$,
$\tilde{W}$ and $\tilde{G}$ are U(1), SU(2) and SU(3) gauge fermion
fields (bino, wino and gluino), respectively.
SU(2) and SU(3) suffices are omitted in (\ref{softbreaking}) for
simplicity.
We assume that the soft SUSY breaking parameters satisfy simple
relations at the GUT scale:
\begin{eqnarray}
  \left( m_Q^2 \right)_i^j &=& \left( m_D^2 \right)_i^j ~=~
  \left( m_U^2 \right)_i^j ~=~ \left( m_L^2 \right)_i^j ~=~
  \left( m_E^2 \right)_i^j ~\equiv~ m_0^2\ \delta_i^j ~,
  \nonumber\\
  \Delta_1^2 &=& \Delta_2^2 ~=~ m_0^2 ~,
  \nonumber\\
  A_D^{ij} &=& f_{DX}^{ij} A_X m_0 ~, ~~
  A_L^{ij} ~=~ f_{LX}^{ij} A_X m_0 ~, ~~
  A_U^{ij} ~=~ g_{UX}^{ij} A_X m_0 ~,
  \nonumber\\
  M_1 &=& M_2 ~=~ M_3 ~\equiv~ M_{gX} ~,
  \label{boundaryconditions}
\end{eqnarray}
where the suffix ``$X$'' stands for the value at the GUT scale.
The boundary conditions (\ref{boundaryconditions}) are due to the
minimal SUGRA model, where local SUSY is spontaneously broken in the
hidden sector which couples to the observable sector (SUSY-GUT, in the
present case) only gravitationally, and hence universal soft SUSY
breaking terms are induced in the observable sector \cite{BFS-CAN-HLW}.


Below the GUT scale, radiative corrections modify all parameters in
the superpotential (\ref{superpotential}) and the soft SUSY breaking
terms (\ref{softbreaking}), as well as three gauge coupling constants
$g_1$, $g_2$ and $g_3$ for U(1), SU(2) and SU(3), respectively.
The evolution of the parameters are described by the RGEs \cite{BKS}.
According to the radiative SU(2) $\times$ U(1) breaking scenario
\cite{IKKT-APW}, we numerically solve the RGEs down to the electroweak scale
$m_Z$ and evaluate the effective potential for the neutral Higgs
fields:
\begin{eqnarray}
  V({\rm Higgs}) &=& V_0 + V_1 ~,
  \nonumber\\
  V_0 &=&   \left( \mu^2 + \Delta_1^2 \right) |h_1|^2
          + \left( \mu^2 + \Delta_2^2 \right) |h_2|^2
          - \left( B\mu h_1 h_2 + {\rm h.~c.} \right)
  \nonumber\\
    & & + \frac{g_1^2 + g_2^2}{8} \left( |h_1|^2 - |h_2|^2 \right)^2 ~,
  \nonumber\\
  V_1 &=& \frac{1}{64\pi^2}
          \str {\cal M}^4 \left( \log \frac{{\cal M}^2}{m_Z^2} -
                                 \frac{3}{2} \right) ~,
  \label{Higgspotential}
\end{eqnarray}
where $\str$ means the supertrace and ${\cal M}$ includes all
(s)quark and (s)lepton masses.
Then the electroweak symmetry breaking condition
\begin{eqnarray}
  \vev{h_1} &=& v \cos \beta ~, ~~
  \vev{h_2} ~=~ v \sin \beta ~,
  \label{vev}
  \\
  m_Z^2 &=& \frac{g_2^2}{2\cos^2\theta_W} v^2 ~,
  \nonumber
\end{eqnarray}
is imposed.

\section{Flavor mixing in the gluino coupling}
\label{sec:mixing}


In order to discuss the flavor mixing in the gluino coupling, we have
to diagonalize the mass matrices for quarks and squarks.
Throughout the calculation hereafter, we choose the basis in the
generation space for the superfields so that the Yukawa coupling
constants for up-type quarks and leptons should be diagonalized {\em
  at the electroweak scale}.
The Yukawa terms in (\ref{superpotential}) are then written as
\begin{equation}
  W_{\rm Yukawa}(m_Z) =
         \hat{f}_D^{kj}
         \left( V_{\rm KM}^\dagger \right)_k^{~i} Q_i D_j H_1
       + \hat{f}_L^{ij} E_i L_j H_1
       + \hat{g}_U^{ij} Q_i U_j H_2 ~,
  \label{YukawaZ}
\end{equation}
where the notation ``$\hat{~~}$'' stands for a diagonal matrix and
$V_{\rm KM}$ is the Kobayashi-Maskawa matrix.
All eigenvalues of $\hat{f}_D$, $\hat{f}_L$ and $\hat{g}_U$ are taken
to be real positive.
Since this choice of the basis is different from that in the GUT
superpotential (\ref{GUTsuperpotential}), a re-diagonalization of the
Yukawa couplings at the GUT scale is needed in order to find the
unification condition of $f_D$ and $f_L$ and the relation between the
Yukawa coupling constants and the dimension-five coupling constants
$C_{L,R}$.
$W_{\rm Yukawa}$ at the GUT scale is diagonalized with appropriate
unitary matrices $U_Q^{(f)}$, $U_D$, $U_E$, $U_L$, $U_Q^{(g)}$ and
$U_U$:
\begin{subequations}
\begin{eqnarray}
  W_{\rm Yukawa}(M_X) &=& f_{DX}^{ij} Q_i D_j H_1
                        + f_{LX}^{ij} E_i L_j H_1
                        + g_{UX}^{ij} Q_i U_j H_2
  \label{YukawaX}
  \\
    &=& \hat{f}_{DX}^{kl} \left( U_Q^{(f)} \right)_k^{~i}
                          \left( U_D \right)_l^{~j} Q_i D_j H_1
      + \hat{f}_{LX}^{kl} \left( U_E \right)_k^{~i}
                          \left( U_L \right)_l^{~j} E_i L_j H_1
  \nonumber\\
  & & + \hat{g}_{UX}^{kl} \left( U_Q^{(g)} \right)_k^{~i}
                          \left( U_U \right)_l^{~j} Q_i U_j H_2 ~.
\end{eqnarray}
\end{subequations}
The unification condition of $f_D$ and $f_L$ is then written as%
\footnote{The ``unification'' of $\hat{f}_{DX}$ and $\hat{f}_{LX}$,
  with quark masses given in Ref.~\protect\cite{GL}, however, is not
  so satisfactory numerically in the first and the second generations
  as the gauge coupling unification.
  We ignore the difference in the present calculation, since we may
  still have ambiguities of the renormalization effect in the very low
  energy region.}
\begin{equation}
  \hat{f}_{DX}^{ij} = \hat{f}_{LX}^{ij} ~.
  \label{YukawaUnification}
\end{equation}
The matter multiplets are accommodated into the SU(5) multiptets as
\begin{eqnarray}
  \Psi_i &\Leftarrow& \left\{ Q_i
          ,~ \left( U_Q^{(g)\dagger} P^\dagger U_U \right)_i^{~j} U_j
          ,~ \left( U_Q^{(f)\dagger} U_E \right)_i^{~j} E_j \right\} ~,
  \nonumber\\
  \Phi_i &\Leftarrow& \left\{ D_i
          ,~ \left( U_D^\dagger U_L \right)_i^{~j} L_j \right\} ~,
  \label{Embedding}
\end{eqnarray}
and the GUT Yukawa coupling constants in Eq.~(\ref{GUTsuperpotential})
are expressed with those in Eq.~(\ref{YukawaX}) as
\begin{eqnarray}
  f^{ij} &=& f_{DX}^{ij} ~,
  \nonumber\\
  g^{ij} &=& g_{UX}^{ik}
             \left( U_U^\dagger P U_Q^{(g)} \right)_k^{~j} ~,
  \label{GUTYukawa}
\end{eqnarray}
where $P$ is a diagonal phase matrix which cannot be absorbed by field
redefinitions in the colored Higgs coupling \cite{EGN}.


The origin of the flavor mixing in the gluino coupling lies in the
difference between the mass basis for quarks and that for squarks.
The mass matrix for up-type squarks is expressed as
\begin{eqnarray}
  -{\cal L}(\mbox{s-up mass}) &=&
       ( \tilde{q}_{u} ,~ \tilde{u}^{\dagger} )
       {\cal M}_{\tilde{u}}^2
        \left( \begin{array}{c}
                 \tilde{q}_{u}^{\dagger} \\ \tilde{u}
               \end{array}\right) ~,
  \nonumber\\
  &=&  ( \tilde{q}_{ui} ,~ \tilde{u}^{\dagger i} )
       \left( \begin{array}{cc}
                \left( m_{LL}^2 \right)^i_j &
                \left( m_{LR}^2 \right)^{ij} \\
                \left( m_{RL}^2 \right)_{ij} &
                \left( m_{RR}^2 \right)_i^j
              \end{array}\right)
              \left( \begin{array}{c}
                       \tilde{q}_{u}^{\dagger j} \\ \tilde{u}_j
                     \end{array}\right) ~,
  \nonumber\\
  \left( m_{LL}^2 \right)^i_j &=&  \left( M_U M_U^\dagger \right)^i_j
    + \left( m_Q^2 \right)^i_j
    + m_W^2 \cos 2\beta \left(   \frac{1}{2}
                               - \frac{1}{6}\tan^2 \theta_W \right)
      \delta^i_j ~,
  \nonumber\\
  \left( m_{RR}^2 \right)_i^j &=& \left( M_U^\dagger M_U \right)_i^j
    + \left( m_U^2 \right)_i^j
    + m_W^2 \cos 2\beta \left( \frac{2}{3}\tan^2 \theta_W \right)
      \delta_i^j   ~,
  \nonumber\\
  \left( m_{LR}^2 \right)^{ij} &=& \mu M_U^{ij} \cot\beta
    + A_U^{ij} v \sin\beta ~,
  \nonumber \\
  m_{RL} &=& m_{LR}^{\dagger} ~,
  \label{squarkmass}
\end{eqnarray}
where $M_U$ is the up-type quark mass matrix $M_U^{ij} = g_U^{ij} v
\sin\beta$ and $\tilde{q}_u$ is the up-type component of the SU(2)
doublet $\tilde{q}$.
The squark mass matrix ${\cal M}_{\tilde{u}}^2$ is not diagonalized
with the quark mass basis (\ref{YukawaZ}) since off-diagonal elements
are induced in the soft SUSY breaking parameter matrices due to the
renormalization effect.
Squark mass basis is obtained by diagonalizing (\ref{squarkmass}) with
a 6$\times$6 unitary matrix $\tilde{U}_U$:
\begin{eqnarray}
  \tilde{u}'_I &=& \left( \tilde{U}_U \right)_I^J \tilde{u}_J ~,
  ~~ I ~=~ 1,\ 2,\ \cdots,\ 6 ~,
  \nonumber\\
  \tilde{u}_I &=&
    \left\{ \begin{array}{lcl}
               \tilde{q}_{uI} & \mbox{for} & I = 1,\ 2,\ 3 \\
               \tilde{u}_{I-3} & \mbox{for} & I = 4,\ 5,\ 6
            \end{array} \right. ~,
  \nonumber\\
  \tilde{U}_U^\dagger {\cal M}_{\tilde{u}}^2 \tilde{U}_U &=&
     \mbox{diagonal} ~,
  \label{DiagonalizeSquark}
\end{eqnarray}
where $\tilde{u}'_I$ is the mass eigenstate of up-type squark.
We define the numbering of $\tilde{u}'_I$ such that the mixing of
$\tilde{u}_I$ is the largest in $\tilde{u}'_I$.
Accordingly we call $\tilde{u}'_1$, $\tilde{u}'_2$, $\cdots$,
$\tilde{u}'_6$ as $\tilde{u}_L$, $\tilde{c}_L$, $\tilde{t}_L$,
$\tilde{u}_R$, $\tilde{c}_R$ and $\tilde{t}_R$ respectively in the
later discussions.
The mass bases of down-type squarks and charged sleptons are obtained
in the same way with 6$\times$6 unitary matrices $\tilde{U}_D$ and
$\tilde{U}_E$, respectively.
Notice that no generation mixing occurs in the lepton/slepton sector
since the right-handed (s)neutrino does not exist in the minimal
model; the nonvanishing off-diagonal elements of the slepton mass
matrix are left-right mixing components only.
Consequently, quark-squark-gluino coupling is written as
\begin{eqnarray}
  {\cal L}_{\rm int}(\mbox{gluino}) &=&
    -i \sqrt{2} g_3 \left\{ \tilde{d}'^{\dagger I}
                            \left(
                            \left( \tilde{U}_D \right)_I^k
                            \left( V_{\rm KM} \right)_k^j
                            \tilde{G} d_{Lj} +
                            \left( \tilde{U}_D \right)_I^{j+3}
                            \ol{\tilde{G}} \ol{d}_{Rj}
                            \right)
                    \right.
  \nonumber\\
    & & ~~~~~~~~~ + \left.  \tilde{u}'^{\dagger I}
                            \left(
                            \left( \tilde{U}_U \right)_I^j
                            \tilde{G} u_{Lj} +
                            \left( \tilde{U}_U \right)_I^{j+3}
                            \ol{\tilde{G}} \ol{u}_{Rj}
                            \right)
                   \right\} + {\rm h.~c.}
  \nonumber\\
    &=&
    -i \sqrt{2} g_3 \left\{ \tilde{d}'^{\dagger I}
                            \left(
                            \left( \tilde{U}'_D \right)_I^j
                            \tilde{G} d_{Lj} +
                            \left( \tilde{U}_D \right)_I^{j+3}
                            \ol{\tilde{G}} \ol{d}_{Rj}
                            \right)
                    \right.
  \nonumber\\
    & & ~~~~~~~~~ + \left.  \tilde{u}'^{\dagger I}
                            \left(
                            \left( \tilde{U}_U \right)_I^j
                            \tilde{G} u_{Lj} +
                            \left( \tilde{U}_U \right)_I^{j+3}
                            \ol{\tilde{G}} \ol{u}_{Rj}
                            \right)
                   \right\} + {\rm h.~c.} ~,
  \label{GluinoCoupling}
\end{eqnarray}
with the definition of $\tilde{U}'_D$ as
\begin{equation}
  \left( \tilde{U}'_D \right)_I^j \equiv
    \left( \tilde{U}_D \right)_I^k \left( V_{\rm KM} \right)_k^j ~.
  \label{primedU}
\end{equation}
$u_{Li}$ and $d_{Li}$ in (\ref{GluinoCoupling}) are left-handed quarks
of mass eigenstates which compose the SU(2) doublet as
\begin{equation}
  Q_i \ni \left( \begin{array}{c}
                   u_{Li} \\
                   \left( V_{\rm KM} \right)_i^j d_{Lj}
                 \end{array}
          \right) ~,
  \label{Doublet}
\end{equation}
and $u_{Ri}$ and $d_{Ri}$ are the fermion components of $U_i$ and
$D_i$, respectively.
Similar flavor mixing formulae are obtained for other gaugino (wino
and bino) coupling terms.

\section{Nucleon decay with dimension five operators}
\label{sec:decay}

As mentioned in Sec.~\ref{sec:model}, the nucleon decay amplitude in
the minimal SU(5) SUSY-GUT model is dominated by the dimension five
operators \cite{SY-W} induced by colored higgsino/Higgs exchanges.
Since the dimension five operators are made from two fermion
(quark/lepton) and two boson (squark/slepton) component fields,
effective baryon number violating four-fermion operators are generated
by one loop ``dressing'' diagrams which involve gauginos or higgsinos
(see Fig.~\ref{fig:diagrams}).
In the present calculation, only the gluino dressing and the charged
wino dressing diagrams are included;
contributions from higgsino dressing diagrams are negligibly small due
to the small Yukawa couplings of light quarks ($u,d,s$), compared to
the SU(2) gauge coupling $g_2$;
neutral wino and bino coupling have the same flavor mixing structure
as that in the gluino coupling, hence their contributions are smaller
than that from the gluino dressing.

The dimension five coupling constants $C_L$ and $C_R$ of
(\ref{superpotential}) at the GUT scale are written in terms of the
Yukawa coupling constants (see (\ref{Embedding}) and
(\ref{GUTYukawa})):
\begin{eqnarray}
  C_{LX}^{ijkl} &=& f_{DX}^{im} \left( U_D^\dagger U_L \right)_m^j
                    g_{UX}^{kn} \left( U_U^\dagger P
                                       U_Q^{(g)} \right)_n^l ~,
  \nonumber\\
  C_{RX}^{ijkl} &=& f_{DX}^{mj} \left( U_Q^{(g)\dagger} P^\dagger
                                       U_U \right)_m^i
                    g_{UX}^{nl} \left( U_Q^{(f)\dagger}
                                       U_E \right)_n^k ~.
  \label{GUTDim5}
\end{eqnarray}
Note that this relation with the index ``$X$'' removed does not hold
true at the electroweak scale.
The effective baryon number violating four-fermion operators at the
electroweak scale are written as
\begin{eqnarray}
  {\cal L}_{\rm eff} (\Delta B = \pm 1) &=&
        \left(   \tilde{C}_{\nu}^{ijkl}(\tilde{G})
               + \tilde{C}_{\nu}^{ijkl}(\tilde{W}) \right)
        \epsilon_{abc} (u_{Lk}^a d_{Ll}^b) (d_{Li}^c \nu_{Lj})
  \nonumber\\
  & & + \left(   \tilde{C}_{e}^{ijkl}(\tilde{G})
               + \tilde{C}_{e}^{ijkl}(\tilde{W}) \right)
        \epsilon_{abc} (u_{Lk}^a d_{Ll}^b) (u_{Li}^c e_{Lj})
  \nonumber\\
  & & + (\mbox{right-handed quark/lepton}) + \mbox{h.~c.} ~,
  \label{4Fermi}
\end{eqnarray}
where $\tilde{C}_{\nu,e}$ are calculated as follows with use of the
numerical values of $C_L$ and $C_R$ at the electroweak scale to be obtained
through their RGEs, squark/slepton mass eigenvalues and the mixing
matrices in the gaugino couplings%
\footnote{Contributions from $C_R$'s and higgsino/neutralino dressings
  are estimated to be small and neglected in the present calculations.
  }:
\begin{eqnarray}
  \tilde{C}_{\nu}^{ijkl}(\tilde{G}) &=& -\frac{4 g_3^2}{3M_C}
    \left\{   \left( C_L^{ijmn} - C_L^{njmi} \right)
              \left( \tilde{U}_U^\dagger \right)_m^I
              \left( \tilde{U}_D^\dagger \right)_n^J
              F_{\tilde{G}}( \tilde{u}'_I, \tilde{d}'_J )
              \left( \tilde{U}_U \right)_I^k
              \left( \tilde{U}'_D \right)_J^l
    \right.
  \nonumber\\
  & & ~~~~~~~~
    \left. - \left( C_L^{mjkn} - C_L^{njkm} \right)
             \left( \tilde{U}_D^\dagger \right)_m^I
             \left( \tilde{U}_D^\dagger \right)_n^J
             F_{\tilde{G}}( \tilde{d}'_I, \tilde{d}'_J )
             \left( \tilde{U}'_D \right)_I^i
             \left( \tilde{U}'_D \right)_J^l
    \right\} ~,
  \nonumber\\
  \tilde{C}_{\nu}^{ijkl}(\tilde{W}) &=& -\frac{g_2^2}{M_C}
    \left\{  \left( C_L^{ijmn} - C_L^{njmi} \right)
             \left( \tilde{U}_U^\dagger \right)_m^I
             \left( \tilde{U}_D^\dagger \right)_n^J
             F_{\tilde{W}}( \tilde{u}'_I, \tilde{d}'_J )
             \left( \tilde{U}'_U \right)_I^k
             \left( \tilde{U}_D \right)_J^l
    \right.
  \nonumber\\
  & & ~~~~~~~~
    \left. + \left( C_L^{mnkl} - C_L^{knml} \right)
             \left( \tilde{U}_U^\dagger \right)_m^I
             \left( \tilde{U}_E^\dagger \right)_n^J
             F_{\tilde{W}}( \tilde{u}'_I, \tilde{e}'_J )
             \left( \tilde{U}'_U \right)_I^i
             \left( \tilde{U}_E \right)_J^j
    \right\} ~,
  \nonumber\\
  \tilde{C}_{e}^{ijkl}(\tilde{G}) &=& -\frac{4 g_3^2}{3M_C}
    \left\{  \left( C_L^{ijmn} - C_L^{mjin} \right)
             \left( \tilde{U}_U^\dagger \right)_m^I
             \left( \tilde{U}_D^\dagger \right)_n^J
             F_{\tilde{G}}( \tilde{u}'_I, \tilde{d}'_J )
             \left( \tilde{U}_U \right)_I^k
             \left( \tilde{U}'_D \right)_J^l
    \right.
  \nonumber\\
  & & ~~~~~~~~
    \left. - \left( C_L^{mjnl} - C_L^{njml} \right)
             \left( \tilde{U}_U^\dagger \right)_m^I
             \left( \tilde{U}_U^\dagger \right)_n^J
             F_{\tilde{G}}( \tilde{u}'_I, \tilde{u}'_J )
             \left( \tilde{U}_U \right)_I^i
             \left( \tilde{U}_U \right)_J^k
    \right\} ~,
  \nonumber\\
  \tilde{C}_{e}^{ijkl}(\tilde{W}) &=& -\frac{g_2^2}{M_C}
    \left\{  \left( C_L^{ijmn} - C_L^{mjin} \right)
             \left( \tilde{U}_U^\dagger \right)_m^I
             \left( \tilde{U}_D^\dagger \right)_n^J
             F_{\tilde{W}}( \tilde{u}'_I, \tilde{d}'_J )
             \left( \tilde{U}'_U \right)_I^l
             \left( \tilde{U}_D \right)_J^k
    \right.
  \nonumber\\
  & & ~~~~~~~~
    \left. + \left( C_L^{mjkl} - C_L^{ljkm} \right)
             \left( \tilde{U}_D^\dagger \right)_m^I
             F_{\tilde{W}}( \tilde{d}'_I, \tilde{\nu}'_j )
             \left( \tilde{U}_D \right)_I^i
    \right\} ~.
  \label{4FermiCoupling}
\end{eqnarray}
Here, $( \tilde{U}'_D )_I^i$ is defined in (\ref{primedU}) and
$( \tilde{U}'_U )_I^i = ( \tilde{U}_U )_I^j (V_{\rm KM})_j^i$.
$F_{\tilde{G}}$ and $F_{\tilde{W}}$ are obtained by the loop integral
\cite{HMY,MATS,NCA}:
\begin{eqnarray}
  F_{\tilde{G}}( \tilde{f}_1, \tilde{f}_2 ) &=&
    \tilde{F}( m_{\tilde{f}_1}, m_{\tilde{f}_2}; M_3 ) ~,
  \nonumber\\
  F_{\tilde{W}}( \tilde{f}_1, \tilde{f}_2 ) &=&
    \left( U_{-} \right)_1^\alpha
    \tilde{F}( m_{\tilde{f}_1}, m_{\tilde{f}_2}; M_{\pm}^\alpha )
    \left( U_{+}^{\dagger} \right)_\alpha^1 ~,
  \label{Floop}
  \\
    \tilde{F}( m_1, m_2; M ) &=&
    \frac{1}{16\pi^2} \frac{M}{m_1^2 - m_2^2}
    \left(   \frac{m_1^2}{m_1^2 - M^2}\log\frac{m_1^2}{M^2}
           - \frac{m_2^2}{m_2^2 - M^2}\log\frac{m_2^2}{M^2}
    \right) ~,
  \nonumber
\end{eqnarray}
where $U_{-}$, $U_{+}$ are 2$\times$2 unitary matrices which diagonalize
the chargino mass matrix and $M_{\pm}^\alpha$ ($\alpha = 1, 2$) are its
eigenvalues:
\begin{eqnarray}
  M(\mbox{chargino}) &=& \left( \begin{array}{cc}
                                  M_2 & \sqrt{2} m_W \sin\beta \\
                                  -\sqrt{2} m_W \cos\beta & -\mu
                                \end{array} \right)
  \nonumber\\
  &=& U_{-} \left( \begin{array}{cc}
                     M_{\pm}^1 & 0 \\
                     0 & M_{\pm}^2
                   \end{array} \right) U_{+}^\dagger ~.
  \label{CharginoMass}
\end{eqnarray}
The low energy QCD correction between $m_Z$ and $1 \GeV$ is taken into
account in order to evaluate the four fermion operators in the next
section.
The quark Lagrangian at $\sim 1 \GeV$ is then converted to the hadron
chiral Lagrangian with $\Delta B = \pm 1$ terms \cite{HMY,CD-CWH} with
use of the matrix element
\begin{equation}
  \vev{ 0 | \epsilon_{abc} (d_L^a u_L^b) u_L^c | p } = \beta_p N_L ~,
  \label{Hadron}
\end{equation}
where $N_L$ is a left-handed proton wave function; it enables us to
evaluate the partial lifetimes of the nucleon decay.

\section{Numerical calculations}
\label{sec:calculations}


According to the framework described in the previous sections, we
calculate the flavor (and left-right) mixing in gaugino couplings and
the nucleon partial lifetimes in a five-dimensional parameter space
$\{m_{\rm top},~ \tan\beta,~ m_0,~ M_{gX},~ \tilde{A}_X\}$, where a
dimensionful $A$ parameter is defined as $\tilde{A}_X \equiv A_X m_0$.
Actual calculations are made in the following procedure.
At first, $m_{\rm top}$ and $\tan\beta$ (at the electroweak scale) are fixed.
Using the numerical values of light quark masses and the
Kobayashi-Maskawa mixing angles given in literatures \cite{GL,PDG}
with the above fixed $m_{\rm top}$ and $\tan\beta$, we evaluate the
Yukawa coupling constants at the electroweak scale (\ref{YukawaZ}).
QCD corrections below the electroweak scale for quark masses other than
$m_{\rm top}$ are included at the one-loop level.
Next, the RGEs for the dimensionless parameters \ie, the gauge coupling
constants and the Yukawa coupling constants are solved upward to the
GUT scale with the boundary conditions at the electroweak scale.
At the GUT scale, the Yukawa coupling constants are re-diagonalized to
obtain the boundary conditions for the dimension-five coupling
constants (\ref{GUTDim5}).
Then the RGEs for the soft SUSY breaking parameters and dimension-five
coupling constants are solved downward with the boundary conditions
(\ref{boundaryconditions}) and (\ref{GUTDim5}).
Since the RGEs are linear for the dimensionful parameters, all soft
SUSY breaking parameters at the electroweak scale are written as linear
combinations of the initial parameters $(m_0,~ M_{gX},~ \tilde{A}_X)$
\cite{MATS}:
\begin{eqnarray}
  \tilde{m}^2_I (m_Z) &=& c_{1I} m_0^2 + c_{2I} M_{gX}^2 +
                  c_{3I} \tilde{A}_X^2 + c_{4I} M_{gX} \tilde{A}_X ~,
  \nonumber\\
  \tilde{M}_J (m_Z) &=& d_{1J} M_{gX} + d_{2J} \tilde{A}_X ~,
  \label{LinearCombinations}
\end{eqnarray}
where $\tilde{m}^2_I$ and $\tilde{M}_J$ are collective notations for
the soft SUSY breaking parameters of mass dimension two
($m^2_{Q,D,U,L,E}$, $\Delta^2_{1,2}$) and one ($M_{1,2,3}$,
$A_{D,U,L}$), respectively.
The coefficients $c$'s and $d$'s are implicit functions of the gauge
couplings and the Yukawa couplings and are determined numerically by
solving the RGEs with four cases of boundary conditions $(m_0,~
M_{gX},~ \tilde{A}_X) = (1,~0,~0)$, $(0,~1,~0)$, $(0,~0,~1)$ and
$(0,~1,~1)$.
Once the coefficients are obtained, the values of soft SUSY breaking
parameters at the electroweak scale for given $(m_0,~ M_{gX},~ \tilde{A}_X)$
are evaluated with the formulae (\ref{LinearCombinations}), and it is
easy to scan the three-dimensional parameter space $\{m_0,~ M_{gX},~
\tilde{A}_X\}$ for fixed $m_{\rm top}$ and $\tan\beta$ with this
method.
The next step is to evaluate the remaining two parameters $\mu$ and
$B$ with the electroelectroweak SU(2) $\times$ U(1) symmetry breaking
condition.
The requirement that the minimum of the Higgs potential
(\ref{Higgspotential}) gives the vacuum expectation values (\ref{vev})
leads to
\begin{eqnarray}
  \mu^2 &=& \frac{\Delta_2^2 - \Delta_1^2}{2\cos 2\beta}
          - \frac{\Delta_1^2 + \Delta_2^2}{2} - \frac{1}{2}m_Z^2
  \nonumber\\
       && - \frac{1}{v\cos 2\beta} \left.
            \left(   \frac{\partial V_1}{\partial h_1} \cos\beta
                   - \frac{\partial V_1}{\partial h_2} \sin\beta
            \right)
            \right|_{\mathop{}^{\scriptstyle h_1 = v\cos\beta}
                              _{\scriptstyle h_2 = v\sin\beta} } ~,
  \label{minimum1}
  \\
  B\mu &=& \left( \mu^2 + \frac{\Delta_1^2 + \Delta_2^2}{2}
           \right) \sin 2\beta
  \nonumber\\
      && + \frac{1}{v} \left.
           \left(   \frac{\partial V_1}{\partial h_1} \sin\beta
                  + \frac{\partial V_1}{\partial h_2} \cos\beta
           \right)
           \right|_{\mathop{}^{\scriptstyle h_1 = v\cos\beta}
                             _{\scriptstyle h_2 = v\sin\beta} } ~.
  \label{minimum2}
\end{eqnarray}
Notice that the solution of the equation (\ref{minimum1}) for
$\mu$ cannot be written in a simple formula since the one-loop part of
the Higgs potential $V_1$ depends on $\mu$.
We solve (\ref{minimum1}) numerically for both signs of $\mu$ and then
calculate $B$ with (\ref{minimum2}).
Since all the MSSM parameters and the dimension-five coupling
constants at the electroweak scale for a given parameter set $(m_{\rm top},~
\tan\beta,~ m_0,~ M_{gX},~ \tilde{A}_X)$ are thus determined, the mass
spectrum of all superparticles and the mixing matrices in the gaugino
couplings are obtained by diagonalizing their mass matrices.


We investigate the parameter space $\{m_0,~ M_{gX},~ A_X\}$ within the
range $10 \GeV \leq m_0 \leq 10 \TeV$, $10 \GeV \leq M_{gX} \leq 10
\TeV$ and $-5 \leq A_X \leq +5$ for each combination of $m_{\rm top}$
= 120, 150 or 180 GeV and $\tan\beta = 2$, 10, 30 or 50.
For $m_{\rm top}$ = 180 GeV and $\tan\beta = 2$, the top Yukawa
coupling diverges below the GUT scale when solving the RGEs.
For $\tan\beta = 50$, no consistent radiative breaking solution is
found for  any $m_{\rm top}$.
Figs.~\ref{fig:mixingu2} -- \ref{fig:mixings3} are histograms for the
specified off-diagonal elements of the gluino coupling matrices
$\tilde{U}_U$ and $\tilde{U}'_D$ for $m_{\rm top}$ = 150 GeV.
The magnitudes of the generation mixing in the right-right and
left-right sectors are small compared to the corresponding left-left
elements.
For a small $\tan\beta$, the left-left elements of $\tilde{U}'_D$ are
approximately equal to the Kobayashi-Maskawa matrix elements:
$(\tilde{U}'_D)_2^1 \approx V_{cd}$, $(\tilde{U}'_D)_3^1 \approx
V_{td}$, \etc in most of the parameter space, while the corresponding
off-diagonal elements of $\tilde{U}_U$ are small: the mass matrices of
up-type quarks, up-type squarks and down-type squarks are diagonalized
in the same basis.
This agrees with the conclusion of Ref.~\cite{BKS} where those mixing
matrices are semi-analytically obtained with an assumption that the
bottom Yukawa coupling is much smaller than the top Yukawa coupling,
which is applicable for small $\tan\beta$.
On the other hand, for a large $\tan\beta$, we find that nonvanishing
off-diagonal elements in the left-left sector of $\tilde{U}_U$ arises
with the same order of magnitudes as $V_{\rm KM}$, which contributes
significantly to the charged lepton decay modes (see below).
Off-diagonal elements of $\tilde{U}'_D$ are smaller than those for
small $\tan\beta$.
The qualitative behavior of the mixing matrices is rather independent
of the different values of $m_{\rm top}$.


Using the obtained values of superpartner masses and mixing matrices,
we evaluate $\tilde{C}$'s in (\ref{4Fermi}) with the formula
(\ref{4FermiCoupling}).
We then take into account of the low energy QCD correction to
$\tilde{C}$'s at the one-loop level%
\footnote{This corresponds to taking $A_{\rm L} \approx 0.46$, where
  $A_{\rm L}$ is the low energy QCD factor used in literatures
  \cite{HMY,MATS,NCA,ENR,AN}. It is argued the two-loop analysis
  gives $A_{\rm L}\approx 0.28$ in Ref.~\cite{AN} }.
We take the chiral Lagrangian factors given in Ref.~\cite{HMY} to
derive amplitudes of various decay modes from (\ref{4Fermi}).
A large uncertainty of the nucleon lifetime comes from the numerical
values of $\beta_p$ and $M_C$.
$\beta_p$ is calculated with various methods \cite{GKSMPT-BEHS}, which
give
\begin{displaymath}
  0.003 \mbox{GeV}^3 \leq \beta_p \leq 0.03 \mbox{GeV}^3 ~,
\end{displaymath}
and Ref.~\cite{HMY,HY} shows
\begin{displaymath}
  2.2 \times 10^{13} \mbox{GeV} \leq M_C \leq
  2.3 \times 10^{17} \mbox{GeV} ~.
\end{displaymath}
Here, we take a small value of $\beta_p$ = 0.003 GeV$^3$ and a large
value of $M_C = 10^{17}$ GeV so that we have a longer nucleon lifetime
for the safety of the later arguments.
Fig.~\ref{fig:lifetime} shows the partial lifetime for each nucleon
decay mode with fixed $m_{\rm top}$ = 150 GeV and $\tan\beta = 2$.
The range of each lifetime comes mainly from the ranges of the soft
SUSY breaking parameters.
As can be seen in the figure, $K \ol{\nu}$ decay modes are dominant
and most severely constrained by the experiments \cite{Kamioka,IMB} as
\begin{displaymath}
  \begin{array}{lcrr}
  \tau( p \rightarrow K^+ \ol{\nu} ) & \geq & 1.0 \times 10^{32} ~
  {\rm yrs} & ~~~~~ {\rm (KAMIOKANDE)} ~, \\
  \tau( p \rightarrow K^+ \ol{\nu} ) & \geq & 6.2 \times 10^{31} ~
  {\rm yrs} & ~~~~~ {\rm (IMB)} ~, \\
  \tau( n \rightarrow K^0 \ol{\nu} ) & \geq & 8.6 \times 10^{31} ~
  {\rm yrs} & ~~~~~ {\rm (KAMIOKANDE)} ~, \\
  \tau( n \rightarrow K^0 \ol{\nu} ) & \geq & 1.5 \times 10^{31} ~
  {\rm yrs} & ~~~~~ {\rm (IMB)} ~.
  \end{array}
\end{displaymath}
Our main concern is to study the contribution of the gluino dressing
diagrams.
To do that, we compare $\tau$(wino) with $\tau$(total), where
$\tau$(wino) is a partial lifetime calculated
with only the wino dressing diagrams taken into account and
$\tau$(total) is that calculated with both wino and gluino dressing
diagrams combined.
The results for $m_{\rm top}$ = 150 GeV are presented in
Figs.~\ref{fig:ratio02} -- \ref{fig:ratio30}.
The nucleon decay amplitude is dominated by the wino dressing diagrams
if $\tau (\mbox{wino}) / \tau(\mbox{total}) \approx 1$, while it is
dominated by the gluino dressing diagrams if $\tau (\mbox{wino}) /
\tau(\mbox{total}) \gg 1$.
There occur cancellations between the wino dressing contributions and
gluino dressing contributions in $\tau (\mbox{wino}) /
\tau(\mbox{total}) \ll 1$ region.
The gluino contributions are small for any modes in the small
$\tan\beta$ case.
For large $\tan\beta$, however, gluino dominant region is realized in
the charged lepton modes.
This is brought about by the following two reasons:
(1) in the simplified analyses based on an assumption of the diagonal
gluino coupling, the gluino dressing diagrams contribute only to the
$K \ol{\nu}$ modes due to the color antisymmetry in the dimension-five
coupling \cite{NCA,ENR}.
In the present case, the gluino coupling is not diagonal any more and
hence the gluino dressing processes contribute to the decay modes
other than $K \ol{\nu}$ modes.
The nonvanishing off-diagonal gluino coupling in the {\em up sector\ }
for large $\tan\beta$ significantly contributes to the charged lepton
modes.
(2) furthermore, if the squark masses are much larger than the gaugino
masses, the amplitude from the gluino dressing diagrams is enhanced by
a factor of $(\alpha_3(m_Z)/\alpha_2(m_Z))^2$ compared to that from the
wino dressing diagrams, since $\tilde{F}$ in Eq.~(\ref{Floop}) is
asymptotically
\begin{equation}
  \tilde{F}(m,m;M) \sim \frac{1}{16\pi^2} \frac{M}{m^2}
  ~~~~~ {\rm for} ~ m \gg M ~,
  \label{asymptoticF}
\end{equation}
and $M_3(m_Z)/M_2(m_Z) = \alpha_3(m_Z)/\alpha_2(m_Z)$ because of the
GUT relation of gaugino masses.
Since (\ref{asymptoticF}) gives an overall suppression of the nucleon
decay amplitude for $m \gg M$ (see Fig.~\ref{fig:md1-M2}), the
dominant gluino contribution in Figs.~\ref{fig:ratio10} and
\ref{fig:ratio30} is realized in the long lifetime region (see
Fig.~\ref{fig:lifetimevsratio}).


Translating the scanned parameters $(m_0,~ M_{gX},~ A_X)$ into the
MSSM parameters $(m_{\tilde{d}_L},~ M_2,~ \mu)$ where we take
$m_{\tilde{d}_L}$, the down-type squark mass of the first generation
as the typical squark mass, we plot the calculated points in the MSSM
parameter space for $m_{\rm top}$ = 150 GeV and $\tan\beta = 2$ in
Figs.~\ref{fig:mu-M2} and \ref{fig:md1-M2}.
The region A in Fig.~\ref{fig:mu-M2} is excluded by LEP constraints on
charginos and neutralinos \cite{LEP}:
\begin{eqnarray}
  m_{\chi^\pm} &>& 45 \GeV ~,
  \nonumber\\
  \Gamma ( Z \rightarrow \chi \chi ) &<& 22 \MeV ~,
  \nonumber\\
  B ( Z \rightarrow \chi \chi' ) &<& 5 \times 10^{-5} ~,
  \nonumber\\
  B ( Z \rightarrow \chi' \chi' ) &<& 5 \times 10^{-5} ~,
  \nonumber
  \label{lep}
\end{eqnarray}
where $\chi^\pm$ is a chargino, $\chi$ is the lightest neutralino and
$\chi'$ is a heavier neutralino.
No solution with radiative SU(2)$\times$U(1) breaking is found in
region B, which is a forbidden region.
Points plotted with small dots are excluded due to the present lower
bound for the proton lifetime $\tau( p \rightarrow K^+ \ol{\nu} ) >
10^{32}$ yrs, giving the constraint of $|\mu| \gsim$ 300 GeV.
This constraint for $\mu$ is roughly unchanged for different $m_{\rm
  top}$ and/or $\tan\beta$.
Fig.~\ref{fig:md1-M2} shows the squark mass bound $m_{\tilde{d}_L}
\gsim 400 \GeV$.
If the lower bound for the proton lifetime is raised to $\tau( p
\rightarrow K^+ \ol{\nu} ) > 10^{33}$ yrs with the near future
experiment at Super-KAMIOKANDE, most of the parameter region with
$m_{\tilde{d}_L} \lsim$ 1 TeV will be excluded.
Lower bounds for other first and second generation squark masses are
found similar to that for $m_{\tilde{d}_L}$, while the bound for the
third generation squarks is lower in general due to the
renormalization effect and the left-right mixing.
Since the nucleon lifetime is approximately proportional to
$(\tan\beta)^{-2}$ \cite{HMY}, the lower bound for the squark mass is
raised for larger $\tan\beta$.
In fact, $\tau( p \rightarrow K^+ \ol{\nu} ) > 10^{32}$ yrs implies
$m_{\tilde{d}_L} \gsim 1 \TeV$ for $\tan\beta = 30$.

\section{Conclusion}
\label{sec:conclusion}

In this paper we have made a systematic analysis of the flavor mixing
in the gaugino couplings within the framework of the minimal SUGRA-GUT.
We have solved the one-loop RGEs for all MSSM parameters including
off-diagonal Higgs coupling matrices with five input parameters,
namely $(m_{\rm top},~ \tan\beta)$ at the electroweak scale $m_Z$ and
$(m_0,~ M_{gX},~ A_X)$ at the GUT scale $M_X$, and we have numerically
obtained full mass spectra and mixing matrices, which satisfy the
radiative electroweak symmetry breaking condition.
For a small $\tan\beta$ ($\tan\beta = 2$), we have obtained a result
consistent with the semi-analytic study \cite{BKS}, in which the top
Yukawa coupling is assumed to be much larger than other Yukawa
couplings: the left-left sector of the generation mixing matrix in the
down-type quark-squark-gluino coupling is approximately equal to the
Kobayashi-Maskawa matrix, while the off-diagonal mixing matrix
elements in the up-type gluino coupling are small.
On the other hand, for large $\tan\beta = 10$ and 30 where the bottom
(and tau, for extremely large $\tan\beta$) Yukawa coupling is not
negligibly small compared with the top Yukawa coupling, we have found
that nonvanishing generation mixing in the up-type gluino coupling
occurs with the magnitudes comparable to the corresponding
Kobayashi-Maskawa matrix elements.
The generation mixing in the down-type gluino coupling is also changed
considerably.

We have applied the generation mixing to the calculation of nucleon
decay widths to study the contributions from the gluino dressing
diagrams compared with the wino dressing diagrams.
In result, it is found that the gluino dressing diagrams give the
dominant contribution to the decay mode containing a charged lepton if
$\tan\beta \gg 1$ and $M_3 \ll m_{\tilde{q}}$ (typical squark mass).
For the charged lepton modes with small $\tan\beta$, or the (anti-)
neutrino emission modes with any $\tan\beta$, the gluino contributions
are relatively small.
In those cases, the contributions from the gluino dressing are at most
of the same order of magnitude as the wino dressing contributions.
We have scanned the MSSM parameter space to find allowed regions with
the present constraints given by the nucleon decay experiments and the
accelerator experiments%
\footnote{In addition, cosmological constraint will be given by the
  analyses of the relic abundance of the lightest superprticle
  \protect\cite{MY-DN-KM}.}.
The latter excludes the parameter region of small superpartner masses,
and the former gives a strict bound to the masses of first and second
generation squarks.
We argue that the whole parameter region with $m_{\tilde{d}_L} \lsim 1
\TeV$ in the minimal SU(5) SUGRA-GUT model can be tested by
Super-KAMIOKANDE.

Our method of calculations and the numerical result itself are
adaptable to the analyses of FCNC in the minimal SUGRA model, which
will be discussed elsewhere.

\subsection*{Acknowledgment}

One of the authors (T.~G.) would like to thank J.~Hisano for helpful
discussions.



\newpage

\section*{Figure Captions}

\newcommand{\diagramsCaption}%
{Examples of dressing diagrams which contribute to the proton decay
  process $p \rightarrow K^+ ~ \ol{\nu}$.}
\newcommand{\mixinguiiCaption}%
{Histograms of the mixing matrix element $(\tilde{U}_U)_2^1$ ($u_L -
  \tilde{c}_L$ mixing) for $\tan\beta = 2$, 10 and 30 with $m_{\rm
    top} = 150 \GeV$, $10 \GeV \leq m_0 \leq 10 \TeV$, $10 \GeV \leq
  M_{gX} \leq 10 \TeV$ and $-5 \leq A_X \leq +5$.}
\newcommand{\mixinguiiiCaption}%
{Histograms of the mixing matrix element $(\tilde{U}_U)_3^1$ ($u_L -
  \tilde{t}_L$ mixing) for $\tan\beta = 2$, 10 and 30 with $m_{\rm
    top} = 150 \GeV$. The parameters are same as those in
  Fig.~\protect\ref{fig:mixingu2}.}
\newcommand{\mixingdiiCaption}%
{Histograms of the mixing matrix element $(\tilde{U}'_D)_2^1$ ($d_L -
  \tilde{s}_L$ mixing) for $\tan\beta = 2$, 10 and 30 with $m_{\rm
    top} = 150 \GeV$. The parameters are same as those in
  Fig.~\protect\ref{fig:mixingu2}.}
\newcommand{\mixingdiiiCaption}%
{Histograms of the mixing matrix element $(\tilde{U}'_D)_3^1$ ($d_L -
  \tilde{b}_L$ mixing) for $\tan\beta = 2$, 10 and 30 with $m_{\rm
    top} = 150 \GeV$. The parameters are same as those in
  Fig.~\protect\ref{fig:mixingu2}.}
\newcommand{\mixingsiCaption}%
{Histograms of the mixing matrix element $(\tilde{U}'_D)_1^2$ ($s_L -
  \tilde{d}_L$ mixing) for $\tan\beta = 2$, 10 and 30 with $m_{\rm
    top} = 150 \GeV$. The parameters are same as those in
  Fig.~\protect\ref{fig:mixingu2}.}
\newcommand{\mixingsiiiCaption}%
{Histograms of the mixing matrix element $(\tilde{U}'_D)_3^2$ ($s_L -
  \tilde{b}_L$ mixing) for $\tan\beta = 2$, 10 and 30 with $m_{\rm
    top} = 150 \GeV$. The parameters are same as those in
  Fig.~\protect\ref{fig:mixingu2}.}
\newcommand{\lifetimeCaption}%
{Nucleon decay partial lifetimes for $m_{\rm top} = 150 \GeV$ and
  $\tan\beta = 2$ with $10 \GeV \leq m_0 \leq 10 \TeV$, $10 \GeV \leq
  M_{gX} \leq 10 \TeV$ and $-5 \leq A_X \leq +5$.
  The lepton in each mode has a lepton number $-1$ (``$\nu$'', ``$e$''
  and ``$\mu$'' mean $\ol{\nu}$, $e^+$ and $\mu^+$, respectively).
  ``$K$'' means a $K$ meson with a $\ol{s}$ quark ($K^+$ or $K^0$).
  $\ol{\nu}$ without a suffix means the total of three neutrinos.
  The shaded region is excluded experimentally.
  If the data points with $\tau(n \rightarrow K^0 \ol{\nu}) < 0.86
  \times 10^{32}$ yrs are omitted, the minimum value of each mode is
  raised to the vertical line.}
\newcommand{\ratioiiCaption}%
{The ratios of partial lifetimes calculated only the wino dressing
  diagrams ($\tau$(wino)) and those calculated with both wino and
  gluino dressing diagrams ($\tau$(total)) for $m_{\rm top}$ = 150 GeV
  and $\tan\beta = 2$ with $10 \GeV \leq m_0 \leq 10 \TeV$, $10 \GeV
  \leq M_{gX} \leq 10 \TeV$ and $-5 \leq A_X \leq +5$.}
\newcommand{\ratioxCaption}%
{The ratios of partial lifetimes calculated only the wino dressing
  diagrams and those calculated with both wino and gluino dressing
  diagrams for $m_{\rm top}$ = 150 GeV and $\tan\beta = 10$ with $10
  \GeV \leq m_0 \leq 10 \TeV$, $10 \GeV \leq M_{gX} \leq 10 \TeV$ and
  $-5 \leq A_X \leq +5$.}
\newcommand{\ratioxxxCaption}%
{The ratios of partial lifetimes calculated only the wino dressing
  diagrams and those calculated with both wino and gluino dressing
  diagrams for $m_{\rm top}$ = 150 GeV and $\tan\beta = 30$ with $10
  \GeV \leq m_0 \leq 10 \TeV$, $10 \GeV \leq M_{gX} \leq 10 \TeV$ and
  $-5 \leq A_X \leq +5$.}
\newcommand{\lifetimevsratioCaption}%
{A scatter plot of the $p \rightarrow K^+ \ol{\nu}$ mode lifetime
  versus the wino-total ratio of the $p \rightarrow K^0 e^+$ mode for
  $m_{\rm top}$ = 150 GeV and $\tan\beta = 10$ with $10 \GeV \leq m_0
  \leq 10 \TeV$, $10 \GeV \leq M_{gX} \leq 10 \TeV$ and $-5 \leq A_X
  \leq +5$. The shaded region is excluded experimentally.}
\newcommand{\muMiiCaption}%
{Scatter plots in $\mu$-$M_2$ plane for $m_{\rm top}$ = 150 GeV and
  $\tan\beta = 2$ with $10 \GeV \leq m_0 \leq 10 \TeV$, $10 \GeV \leq
  M_{gX} \leq 10 \TeV$ and $-5 \leq A_X \leq +5$.
  Region A is excluded by LEP experiment.
  Region B has no radiative breaking solutions (see text).
  The region plotted with small dots should be excluded by the nucleon
  decay experiments.}
\newcommand{\mdiMiiCaption}%
{Scatter plots in $m_{\tilde{d}_L}$-$M_2$ plane for $m_{\rm top}$ =
  150 GeV and $\tan\beta = 2$ with $10 \GeV \leq m_0 \leq 10 \TeV$,
  $10 \GeV \leq M_{gX} \leq 10 \TeV$ and $-5 \leq A_X \leq +5$.
  The region plotted with small dots should be excluded by the nucleon
  decay experiments.}
\begin{list}{\bf FIG.~??}{\relax}
\Fig{fig:diagrams} \diagramsCaption
\Fig{fig:mixingu2} \mixinguiiCaption
\Fig{fig:mixingu3} \mixinguiiiCaption
\Fig{fig:mixingd2} \mixingdiiCaption
\Fig{fig:mixingd3} \mixingdiiiCaption
\Fig{fig:mixings1} \mixingsiCaption
\Fig{fig:mixings3} \mixingsiiiCaption
\Fig{fig:lifetime} \lifetimeCaption
\Fig{fig:ratio02} \ratioiiCaption
\Fig{fig:ratio10} \ratioxCaption
\Fig{fig:ratio30} \ratioxxxCaption
\Fig{fig:lifetimevsratio} \lifetimevsratioCaption
\Fig{fig:mu-M2} \muMiiCaption
\Fig{fig:md1-M2} \mdiMiiCaption
\end{list}

\newpage

\section*{Figures}

\vfil
\begin{figure}[h]
  \begin{center}
    \leavevmode
  \end{center}
  \caption{\diagramsCaption}
  \label{fig:diagrams}
\end{figure}
\begin{subfigures}
\vfil

\begin{figure}[p]
  \begin{center}
    \leavevmode
  \end{center}
  \caption{\mixinguiiCaption}
  \label{fig:mixingu2}
\end{figure}

\begin{figure}[p]
  \begin{center}
    \leavevmode
  \end{center}
  \caption{\mixinguiiiCaption}
  \label{fig:mixingu3}
\end{figure}

\begin{figure}[p]
  \begin{center}
    \leavevmode
  \end{center}
  \caption{\mixingdiiCaption}
  \label{fig:mixingd2}
\end{figure}

\begin{figure}[p]
  \begin{center}
    \leavevmode
  \end{center}
  \caption{\mixingdiiiCaption}
  \label{fig:mixingd3}
\end{figure}

\begin{figure}[p]
  \begin{center}
    \leavevmode
  \end{center}
  \caption{\mixingsiCaption}
  \label{fig:mixings1}
\end{figure}

\begin{figure}[p]
  \begin{center}
    \leavevmode
  \end{center}
  \caption{\mixingsiiiCaption}
  \label{fig:mixings3}
\end{figure}

\end{subfigures}

\begin{figure}[p]
  \begin{center}
    \leavevmode
  \end{center}
  \caption{\lifetimeCaption}
  \label{fig:lifetime}
\end{figure}

\begin{subfigures}

\begin{figure}[p]
  \begin{center}
    \leavevmode
  \end{center}
  \caption{\ratioiiCaption}
  \label{fig:ratio02}
\end{figure}

\begin{figure}[p]
  \begin{center}
    \leavevmode
  \end{center}
  \caption{\ratioxCaption}
  \label{fig:ratio10}
\end{figure}

\begin{figure}[p]
  \begin{center}
    \leavevmode
  \end{center}
  \caption{\ratioxxxCaption}
  \label{fig:ratio30}
\end{figure}

\end{subfigures}

\begin{figure}[p]
  \begin{center}
    \leavevmode
    \makebox[0cm]{
      }
  \end{center}
  \caption{\lifetimevsratioCaption}
  \label{fig:lifetimevsratio}
\end{figure}

\begin{subfigures}

\begin{figure}[p]
  \begin{center}
    \makebox[0cm]{
      }
  \end{center}
  \caption{\muMiiCaption}
  \label{fig:mu-M2}
\end{figure}

\begin{figure}[p]
  \begin{center}
    \makebox[0cm]{
      }
  \end{center}
  \caption{\mdiMiiCaption}
  \label{fig:md1-M2}
\end{figure}

\end{subfigures}

\end{document}